\documentclass[aps,prb,article,twocolumn,showpacs,preprintnumbers,amsmath,amssymb,superscriptaddress]{revtex4}
\date{\today}
\usepackage{epsfig}
\usepackage{subfigure}
\usepackage{graphicx}% Include figure files
\usepackage{dcolumn}% Align table columns on decimal point
\usepackage{bm}% bold math
\usepackage[colorlinks,linkcolor=blue,hyperindex,CJKbookmarks]{hyperref}
\usepackage{float}
\usepackage{hyperref}
\usepackage{comment}
\newcommand{\imag}      {\mathrm{Im}}

\newcommand{\Ham}   {{\mathcal{H}}}

\newcommand{\kbf}      {\textbf{k}}
\newcommand{\qbf}      {\textbf{q}}

\newcommand{\ibf}      {\textbf{i}}
\newcommand{\jbf}      {\textbf{j}}
\newcommand{\rbf}      {\textbf{r}}

\begin{document}

\title{Theory of Time-Resolved Raman Scattering in Correlated Systems:\\ Ultrafast Engineering of Spin Dynamics and Detection of Thermalization}
\author{Yao Wang }
\affiliation{Department of Physics, Harvard University, Cambridge, Massachusetts 02138, USA}

\author{Thomas P. Devereaux}
\affiliation{Stanford Institute for Materials and Energy Sciences, SLAC National Accelerator Laboratory, 2575 Sand Hill Road,
Menlo Park, California 94025, USA}
\affiliation{Geballe Laboratory for Advanced Materials, Stanford University, California 94305, USA}

\author{Cheng-Chien Chen}
\affiliation{Department of Physics, University of Alabama at Birmingham, Birmingham, Alabama 35294, USA}

\date{\today}

\begin{abstract}
Ultrafast characterization and control of many-body interactions and elementary excitations are critical to understanding and manipulating emergent phenomena in strongly correlated systems. In particular, spin interaction plays an important role in unconventional superconductivity, but efficient tools for probing spin dynamics especially out of equilibrium, are still lacking. To address this question, we develop a theory for nonresonant time-resolved Raman scattering, which can be a generic and powerful tool for nonequilibrium studies. We also use exact diagonalization to simulate the pump-probe dynamics of correlated electrons in the square-lattice single-band Hubbard model. Different ultrafast processes are shown to exist in the time-resolved Raman spectra and dominate under different pump conditions. For high-frequency and off-resonance pumps, we show that the Floquet theory works well in capturing the softening of bimagnon excitation. By comparing the Stokes/anti-Stokes spectra, we also show that effective heating dominates at small pump fluences, while a coherent many-body effect starts to take over at larger pump amplitudes and frequencies on resonance to the Mott gap. Time-resolved Raman scattering thereby provides the platform to explore different ultrafast processes and design material properties out of equilibrium.
\end{abstract}
\maketitle

\section{Introduction}
Ultrafast detection and engineering of physical properties are the ultimate goal of nonequilibrium studies\cite{zhang2014dynamics, basov2017towards, wang2018theoretical}. Among different degrees of freedom in solids, spin physics plays an important role in unconventional superconductivity\cite{tsuei2000pairing, scalapino2012acommon, maier2016pairing}, frustrated magneitsm\cite{meng2010quantum, balents2010spin}, magnetism materials\cite{chatterji2005neutron}, and spintronics\cite{wolf2001spintronics,vzutic2004spintronics}. Understanding collective spin excitations out of equilibrium is also crucial for the explanation of photoinduced emergent phenomena like transient superconductivity\cite{fausti2011light, mitrano2016possible,wang2017light}. Due to the fluence limitation, however, the spin-sensitive inelastic neutron scattering cannot be applied as an ultrafast technique. Therefore, although nonequilibrium dynamical spin structure factors were predicted theoretically\cite{wang2014real, mentink2015ultrafast, claassen2017dynamical, wang2017light}, they cannot be directly measured in ultrafast experiments. With the recent advance of photon spectroscopies, probing spin dynamics through the charge channel has become promising\cite{batignani2015probing,dean2016ultrafast,cao2018ultrafast}. For example, equilibrium Raman scattering was used to measure bimagnon excitations and provide information for the underlying spin interactions\cite{sugai1988two, singh1989, devereaux2007inelastic, chen2011}. Time-resolved Raman scattering\cite{lee1979time} was employed to detect lattice and molecule vibrations\cite{kash1985subpicosecond, ruhman1988coherent, chesnoy1988resonant, weiner1991femtosecond, kahan2007following, schnedermann2016vibronic, batignani2016electronic, jen2017ultrafast, ferrante2018resonant}, and has been pushed forward to study collective excitations of quantum materials in recent years\cite{batignani2015probing, bowlan2018using}. However, without a microscopic nonequilibrium theory, a systematic and predictable engineering in correlated systems is still not practical to date. As we shall demonstrate theoretically below, time-resolved Raman spectroscopy can provide a platform to distinguish different ultrafast procedures and pave the way to precise engineering of spin interactions out of equilibrium.

\begin{figure}[!t]
\includegraphics[width=\columnwidth]{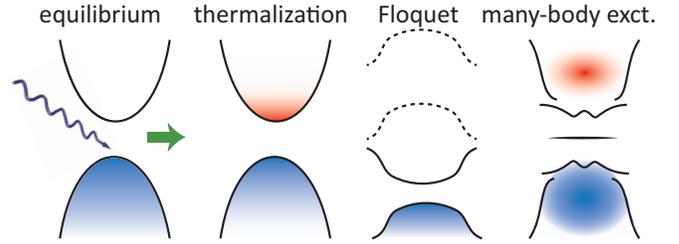}
\caption{\label{fig:cartoon}
Different ultrafast processes induced by a pump field: effective heating or thermalization, transient (Floquet) band renormalization, and non-thermal many-body excitation.
}
\end{figure}

On general grounds, pump-induced ultrafast behaviors include effective heating, transient Floquet band renormalization, and nonthermal many-body excitation. These processes are sketched in Fig.~\ref{fig:cartoon}. Under an infinitely long periodic driving field, a system is known to exhibit a superposition of Floquet steady states\cite{wang2013observation,mahmood2016selective}. Via photoassisted virtual hoppings, the corresponding band renormalization and replicas can correct the effective spin exchange $J$\cite{itin2015effective,mentink2015ultrafast}. Since the Floquet steady states can be precisely predicted by the external pump conditions, this renormalization effect can be adopted to engineer the underlying physical parameters. However, this process requires infinitely long pump and off resonance with direct excitations across a charge gap. This is not practical in realistic experiments, where the pump pulse has a finite-time profile, and the duration of an infrared or terahertz pump can be comparable to its oscillation period. Thus, a resonant excitation is unavoidable due to the existence of higher-energy unoccupied states. This residual resonance to the lowest order can cause effective heating\cite{dalessio2014long,lazarides2014equilibrium, bukov2016schrieffer}, unless the systems are integrable or many-body localized as in ideal theoretical scenarios\cite{dalessio2013many,ponte2015many, lazarides2015fate}. Reducing the pump width and probe delay can suppress thermalization\cite{kuwahara2016floquet, mori2016rigorous, abanin2015exponentially}, but restrict exotic Floquet physics. Moreover, many-body physics can lead to nonlinear modulation of electronic structure, which cannot be attributed simply to effective heating\cite{werner2012nonthermal,moeckel2008interaction, wang2017producing}. The above three effects exist in realistic ultrafast experiments, and their interplay determines the final Raman spectra in the time domain.

In this paper, we derive the theory for nonresonant time-resolved Raman scattering and use exact diagonalization to compute the pump-probe Raman spectra for a square-lattice single-band Hubbard model. We show that bimagnon excitations can be reflected in Raman spectra, and that each mechanism depicted in Fig.~\ref{fig:cartoon} can become dominant under different pump conditions. In particular, a low-frequency resonant pump results in clear thermalization, while a high-frequency nonresonant pump causes a Floquet renormalization of energy-scale $J$. Both scenarios, however, are violated for extremely strong pumps, where many-body excitations take over. Being the first theoretical investigation of time-resolved Raman scattering in strongly correlated materials, our work provides a platform to study different ultrafast mechanisms and nonequilibrium spin excitations. These different mechanisms can be observed over a wide range of pump conditions from infrared to ultraviolet lasers, and our results are expected to be valid for a variety of correlated electron systems.

The rest of this paper is organized as follows. In Sec.~\ref{sec:theory}, we derive the theories of Raman spectroscopies both in and out of equilibrium, with a focus on nonresonant Raman scattering. In Sec.~\ref{sec:result}, we show simulations of time-resolved Raman spectra for a correlated Hubbard system in a pump-probe experiment. Floquet engineering of spin exchange interactions and extraction of effective temperature are also discussed. We conclude the paper in Sec.~\ref{sec:conclusion} by summarizing our main results.

\section{Theory of Time-Resolved Raman Scattering}\label{sec:theory}
The influence of an electromagnetic field for single band systems can be introduced through a Peierls substitution $c_{\mathbf{i}\sigma} \rightarrow c_{\mathbf{i}\sigma} e^{-i\int_{-\infty}^{\textbf{r}_i} \textbf{A}(\rbf^\prime,t)\cdot d\mathbf{r}^\prime} $. Here, $c_{\mathbf{i}\sigma}$ is a fermionic annihilation operator for an electron of spin $\sigma$ on lattice site $\mathbf{i}$, and $\textbf{A}$ is a vector potential containing both pump and probe fields. Since the pump is typically strong and explicitly treated, we denote the pump Hamiltonian as $\Ham_0$ and expand the Hamiltonian $\Ham$ in powers of the probe field $\textbf{A}^{\rm(pr)}$\cite{devereaux2007inelastic}:
\begin{eqnarray}\label{chp2:eleLightHam}
\Ham(t) &=& \Ham_0(t) + \Ham_{\rm pr}(t)\\
\Ham_{\rm pr}(t) &\approx&   -\sum_{\rbf,\alpha} \hat{j}_{\alpha}(\rbf) A_{\alpha}^{\rm(pr)}(\rbf,t)\!\nonumber\\
&&-\frac{1}{2} \sum_{\rbf,\alpha,\beta} \hat{\gamma}_{\alpha\beta}(\rbf)A_{\alpha}^{\rm(pr)}(\rbf,t)^*A_{\beta}^{\rm(pr)}(\rbf,t),
\end{eqnarray}
where $\alpha$ denotes the light polarization direction. For Hamiltonians with nearest-neighbor (N.N.) hopping, $A_{\alpha}^{\rm(pr)}\!(\rbf,t) \!=\! \int_{\rbf}^{\rbf+\mathbf{1}_\alpha}\!\textbf{A}^{\rm(pr)}\!(\rbf^\prime,t)\!\cdot\! d\mathbf{r}^\prime$, the paramagnetic current density operator $\hat{j}_{\alpha}(\rbf_\ibf) = it_h\sum_\sigma (c_{\ibf+\mathbf{1}_\alpha,\sigma}^\dagger c_{\ibf \sigma} - c_{\ibf\sigma}^\dagger c_{\ibf+\mathbf{1}_\alpha, \sigma})$, and the scattering vertex $\hat{\gamma}_{\alpha}(\rbf_\ibf) = -t_h\sum_\sigma(c_{\ibf\sigma}^\dagger c_{\ibf+\mathbf{1}_\alpha, \sigma}+c_{\ibf+\mathbf{1}_\alpha,\sigma}^\dagger c_{\ibf \sigma})$. We consider the whole procedure starting from the equilibrium ground state of the static Hamiltonian $\Ham_0(t= -\infty)$. Therefore, for the equilibrium spectrum, $\Ham_0(t)\equiv \Ham_0$ is time-independent; for the nonequilibrium pump-probe spectrum, $\Ham_0(t)$ contains the original Hamiltonian with the presence of a time-dependent pump field.

In the Fourier space of momentum transfer $\qbf$, the vector potential reads $A_\alpha^{\rm(pr)}\!(\qbf,t)\! =\! \frac1{N}\sum_\rbf e^{-i\qbf\cdot \rbf} A_\alpha^{\rm(pr)}\!(\rbf,t)$. With the effective mass approximation, $\hat{j}_\alpha\! (\qbf)\!=\!\sum_{\kbf\sigma} ({\partial \varepsilon_\kbf }/{\partial k_\alpha}) c_{\kbf+\qbf/2,\sigma}^\dagger\! c_{\kbf-\qbf/2,\sigma}$, and $\hat{\gamma}_{\alpha\beta}\!(\qbf)\!=\!\sum_{\kbf\sigma} ({\partial^2 \varepsilon_\kbf }/{\partial k_\alpha \partial k_\beta})c_{\kbf+\qbf/2,\sigma}^\dagger c_{\kbf-\qbf/2,\sigma}$. The fermionic operator $c_{\kbf\sigma}$ annihilates an electron of momentum $\kbf$ and spin $\sigma$. On a square lattice with N.N. hopping amplitude $t_h$, the band structure $\varepsilon_\kbf$ is $-2t_h(\cos k_x + \cos k_y)$. The probe Hamiltonian then becomes
\begin{eqnarray}
\mathcal{H}_{\rm pr}(t) &=& -\sum_{\qbf,\alpha} \hat{j}_{\alpha}(\qbf) A_{\alpha}^{\rm(pr)}(\qbf,t)\nonumber\\
&&-\frac12 \sum_{\qbf,\qbf_i\atop\alpha,\beta} \hat{\gamma}_{\alpha\beta}(\qbf)A_{\alpha}^{\rm(pr)}(\qbf_s,t)^*A_{\beta}^{\rm(pr)}(\qbf_i,t),
\end{eqnarray} 
where $\qbf_i$ ($\qbf_s$) is the incident (scattering) photon momentum, and $\qbf \equiv \qbf_i-\qbf_s$ is the net momentum transfer.

Within the linear-response theory, the cross sections of various photon spectroscopies can be obtained through a perturbative expansion. Below we first recapitulate the theory of equilibrium Raman spectroscopy\cite{devereaux2007inelastic}. We then derive the nonequilibrium pump-probe Raman cross section, with a focus on nonresonant scattering. We compare both formalisms at the end of this section and discuss a probe-induced linewidth broadening.

\subsection{Equilibrium Raman Cross Section}
While the single-photon absorption ($\propto\!A_\alpha^{\rm(pr)}$) concerns measuring the photocurrent in optical conductivity, Raman scattering as a photon-in-photon-out procedure ($\propto\!A_{\alpha}^{\rm(pr)}\!A_{\beta}^{\rm(pr)}$) probes particle-hole excitations with a form factor. The Raman cross section is proportional to the transition rate determined by Fermi's golden rule:
\begin{eqnarray}\label{chp2:transitionRate}
\mathcal{R}(\qbf, \omega_i,\omega_s) = \sum_{n}\Big|\big\langle n \big| \hat{M}(\qbf, \omega_i,\omega_s) \big| G\big\rangle\Big|^2\delta(\omega+E_G-E_n),
\end{eqnarray}
where $|G\rangle$ and $|n\rangle$ are respectively the ground and excited states, and $\omega = \omega_i -\omega_s$ is the photon energy loss.

The effective light-scattering operator $\hat{M}(\qbf, \omega_i,\omega_s)$ contains the ``resonant'' and ``nonresonant'' processes $\hat{M}(\qbf, \omega_i,\omega_s)= \hat{M}^{\rm R}(\qbf, \omega_i,\omega_s)+ \hat{M}^{\rm N}(\qbf, \omega) $. The resonant scattering operator is
\begin{eqnarray}\label{chp2:scatteringOperatorR}
\hat{M}^{\!\rm R}\!(\qbf, \!\omega_i,\!\omega_s)\! &=&\! \sum_{\alpha,\beta}  
\!\left[-\hat{j}_{\beta}(\qbf_s)^\dagger \frac1{\Ham\!-\!E_G \!-\!\omega_i\!-\!i0_+}\hat{j}_{\alpha}(\qbf_i)\right.\nonumber\\
&&\left.+\hat{j}_{\alpha}\!(\qbf_i)^\dagger\frac1{\Ham\!-\!E_G \!+\!\omega_s\!+\!i0_+}\hat{j}_{\beta}\!(\qbf_s)\right]\hat{\bf e}^{(i)}_\alpha \hat{\bf e}^{(s)}_\beta, \nonumber\\
\end{eqnarray}
which involves processes through resonant intermediate states. The nonresonant scattering operator is
\begin{eqnarray}\label{chp2:scatteringOperatorN}
\hat{M}^{\rm N}(\qbf, \omega)=\sum_{\alpha,\beta} \hat{\bf e}^{(i)}_\alpha \hat{\bf e}^{(s)}_\beta
\hat{\gamma}_{\alpha\beta}(\qbf),
\end{eqnarray}
where $\hat{\bf e}^{(i)}_\alpha$ and $\hat{\bf e}^{(s)}_\beta $ denote respectively the polarizations of incident and scattering photons, which can be configured in experiment. In this sense, $\hat{M}^{\rm N}(\qbf, \omega_i,\omega_s)$ describes a collective excitation with direct energy $\omega$ and momentum $\qbf$ transferred between electrons and photons.

In optical Raman scattering $\qbf\approx0$, the transition into resonant intermediate states can be ignored. The resulting nonresonant Raman response then reads
\begin{eqnarray}\label{chp2:nonResRaman}
{\mathcal{R}^N}(\qbf, \omega)= \frac1{\pi}\imag\left\langle G\left|\hat\gamma^\dagger(\qbf) \frac1{\mathcal{H}-E_G-\omega-i\delta} \hat\gamma(\qbf) \right|G\right\rangle,
\end{eqnarray}
where $\hat\gamma(\qbf) = \sum_{\alpha,\beta} \hat{\bf e}^{(i)}_\alpha  \hat\gamma_{\alpha\beta}(\qbf)\hat{\bf e}^{(s)}_\beta $, and $\delta$ is a phenomenological lifetime broadening effect. Below we focus on the long-wavelength optical limit $\qbf\approx0$ and omit the $\qbf$ label. Here we present the result at zero temperature in order to facilitate the comparison with pure-state dynamics later. A finite-temperature spectrum can be obtained through an ensemble average of Eq.~\eqref{chp2:nonResRaman} over the full Hilbert space.

On a tetragonal lattice, the scattering vertices can be decomposed into the irreducible representation of the $D_{4h}$ point group\cite{devereaux2007inelastic}. Specifically, in the $A_{1g} (xx+yy)$ channel
\begin{eqnarray}\label{eq:eqA1ggamma}
  \hat{\gamma}_{A_{1g}} &=& \sum_{\textbf{k}\sigma}\left(\frac{\partial^2}{\partial k_x^2}+\frac{\partial^2}{\partial k_y^2}\right)\varepsilon_\textbf{k} c_{\textbf{k},\sigma}^\dagger c_{\textbf{k},\sigma},
\end{eqnarray}
and in the $B_{1g} (xx-yy)$ channel
\begin{eqnarray}\label{eq:eqB1ggamma}
  \hat{\gamma}_{B_{1g}} &=& \sum_{\textbf{k}\sigma}\left(\frac{\partial^2}{\partial k_x^2}-\frac{\partial^2}{\partial k_y^2}\right)\varepsilon_\textbf{k} c_{\textbf{k},\sigma}^\dagger c_{\textbf{k},\sigma}.
\end{eqnarray}
With N.N. hopping, the vertices in the brackets are proportional to ($\cos k_x +\cos k_y$) and ($\cos k_x -\cos k_y$), respectively. With only nearest-neighbor hoppings and time-reversal symmetry, the $A_{2g}(xy-yx)$ and $B_{2g}(xy+yx)$ channels both vanish.

\subsection{Nonequilibrium Raman Cross Section}
While one could phenomenologically extend Eq.~\eqref{chp2:transitionRate} to nonequilibrium without a precise treatment of the probe profile, here, we present a detailed derivation by considering explicitly a quantized photon field. By doing so, the derivation can be extended to other nonequilibrium spectroscopies such as time-resolved x-ray absorption and resonant inelastic x-ray scattering, where the creation and annihilation of photons are necessary.

In the second quantization of the photon field, $A_{\alpha}^{\rm(pr)}(\qbf) = a_{\qbf\alpha}+a_{-\qbf\alpha}^\dagger$, and $\Ham_{\rm pr}$ can be rewritten as
\begin{eqnarray}\label{eq:probeHamDiv}
\mathcal{H}_{\rm pr}(t) = \mathcal{H}_{\rm pr}^{\rm (ab)}(t) + \mathcal{H}_{\rm pr}^{\rm (ab2)}(t) + \mathcal{H}_{\rm pr}^{\rm (sc)}(t) +h.c.
\end{eqnarray}
Here, the single-photon absorption part is 
\begin{eqnarray}
	\mathcal{H}_{\rm pr}^{\rm (ab)}(t)  = -\sum_{\qbf,\alpha} \hat{j}_{\alpha}(\qbf,t) a_{\qbf\alpha},
\end{eqnarray}
the two-photon absorption part is 
\begin{eqnarray}
	\mathcal{H}_{\rm pr}^{\rm (ab2)}(t)  = -\frac12 \sum_{\qbf_i,\qbf_s\atop\alpha,\beta} \hat{\gamma}_{\alpha\beta}(\qbf_i-\qbf_s,t)a_{-\qbf_s\alpha}a_{\qbf_i\beta},
\end{eqnarray}
and the scattering part is 
\begin{eqnarray}
	\mathcal{H}_{\rm pr}^{\rm (sc)}(t)  = -\frac12 \sum_{\qbf_s,\qbf_s\atop\alpha,\beta} \hat{\gamma}_{\alpha\beta}(\qbf_i-\qbf_s,t)a_{\qbf_s\alpha}^\dagger a_{\qbf_i\beta}.
\end{eqnarray}
Their Hermitian conjugates are written separately in Eq.~\eqref{eq:probeHamDiv}, and $\mathcal{H}_{\rm pr}^{\rm (sc)}$ itself is Hermitian. In contrast to the equilibrium situation, the pump-probe procedure involves the impact of the pump field in the probe Hamiltonian. Specifically, the fermionic momenta in $\hat j_{\alpha}$ and $\hat\gamma_{\alpha\beta}$ are shifted by the instantaneous pump field $\kbf \rightarrow \kbf - \mathbf{A}(t)$. Therefore, $\mathcal{H}_{\rm pr}(t)$ still has explicit time dependence in the linear-response expansion.

We proceed by expanding the unitary time propagator $\mathcal{U}$ in terms of the probe Hamiltonian $\mathcal{H}_{\rm pr}$ to second order:
\begin{widetext}
\begin{eqnarray}\label{eq:Uexpr}
\mathcal{U}(t,-\infty) &\approx& \mathcal{T}e^{-i\int_{-\infty}^t \!\Ham_0(\tau) d\tau}- i\int_{-\infty}^t\mathcal{U}_0(t,\tau)  \mathcal{H}_{\rm pr}(\tau)  \mathcal{U}_0(\tau,-\infty) d\tau\nonumber\\
&&  - i\int_{-\infty}^t\!dt_2 \int_{-\infty}^{t_2}\!dt_1\,\mathcal{U}_0  (t,t_2)  \mathcal{H}_{\rm pr}(t_2) \mathcal{U}_0(t_2,t_1)  \mathcal{H}_{\rm pr}(t_1)\mathcal{U}_0(t_1,-\infty) \nonumber\\
&=& \mathcal{T}e^{-i\int_{-\infty}^t \!\Ham_0(\tau) d\tau}- i\int_{-\infty}^t\mathcal{U}_0(t,\tau) \mathcal{H}_{\rm pr}^{\rm (ab)}(\tau) \mathcal{U}_0(\tau,-\infty) d\tau- i\int_{-\infty}^t\mathcal{U}_0(t,\tau) \mathcal{H}_{\rm pr}^{\rm (ab2)}(\tau) \mathcal{U}_0(\tau,-\infty) d\tau\nonumber\\
&&    - i\int_{-\infty}^t\!dt_2 \int_{-\infty}^{t_2}\!dt_1\,\mathcal{U}_0  (t,t_2)  \mathcal{H}_{\rm pr}^{\rm (ab)}(t_2) \mathcal{U}_0(t_2,t_1)  \mathcal{H}_{\rm pr}^{\rm (ab)}(t_1)\mathcal{U}_0(t_1,-\infty) \nonumber\\
&& - i\int_{-\infty}^t\!dt_2 \int_{-\infty}^{t_2}\!dt_1\,\mathcal{U}_0  (t,t_2)  \mathcal{H}_{\rm pr}^{\rm (ab)*}(t_2) \mathcal{U}_0(t_2,t_1)  \mathcal{H}_{\rm pr}^{\rm (ab)}(t_1)\mathcal{U}_0(t_1,-\infty) - i\int_{-\infty}^t\mathcal{U}_0(t,\tau) \mathcal{H}_{\rm pr}^{\rm (sc)}(\tau) \mathcal{U}_0(\tau,-\infty) d\tau.\nonumber\\
\end{eqnarray}
\end{widetext}
Here, we denote the unperturbed propagator as
\begin{eqnarray}
	\mathcal{U}_0(t_2,t_1)=\mathcal{T}e^{-i\int_{t_1}^{t_2} \!\Ham_0(\tau) d\tau}.
\end{eqnarray}
Since the equilibrium ground-state wavefunction is usually selected to be $\big| \psi(t = -\infty) \big\rangle$, the Hermitian conjugate terms of $\mathcal{H}_{\rm pr}^{\rm (ab)}$ and $\mathcal{H}_{\rm pr}^{\rm (ab2)}$ do not contribute to the first four integrals, as the ground state cannot emit any photons. The second term in Eq.~\eqref{eq:Uexpr} is the single-photon absorption related to linear optical conductivity. The third and fourth terms are the two-photon absorption reflected in nonlinear conductivity. The last two terms correspond to photon scattering with a conserved photon number. The scattering amplitude and Raman intensity are related to $\mathcal{S}_{\qbf_i\qbf_s}^{\alpha\beta}=a_{\qbf_i\beta}a_{\qbf_i\alpha}^\dagger$ and $\mathcal{O}_{\qbf_i\qbf_s}^{\alpha\beta}=\mathcal{S}_{\qbf_i\qbf_s}^{\alpha\beta\dagger}\mathcal{S}_{\qbf_i\qbf_s}^{\alpha\beta}$, respectively. The photon-in-photon-out scattering operator selectively detects the last two integrals in Eq.~\eqref{eq:Uexpr} through the observable $\langle\mathcal{O}\rangle(t)$:
\begin{widetext}
\begin{eqnarray}\label{eq:crossSec1}
\langle\mathcal{O}\rangle(t) &=& \big\langle \psi(-\!\infty) \big| \mathcal{U}_0(-\infty,t) \mathcal{O}_{\qbf_i\qbf_s}^{\alpha\beta}\mathcal{U}_0(t,-\infty)\big|\psi(-\!\infty) \big\rangle\nonumber\\
 && + 4\iint_{-\infty}^t d\tau d\tau^\prime\big\langle \psi(-\!\infty) \big|\mathcal{U}_0(-\infty, \tau^\prime)  \mathcal{H}_{\rm pr}^{\rm (sc)\dagger}(\tau^\prime)\mathcal{U}_0(\tau^\prime,t)
 a_{\qbf_i\alpha}a^\dagger_{\qbf_s\beta}a_{\qbf_s\beta}a^\dagger_{\qbf_i\alpha}
 \mathcal{U}_0(t,\tau)\mathcal{H}_{\rm pr}^{\rm (sc)}(\tau)\mathcal{U}_0(\tau,-\infty)\big|\psi(-\!\infty) \big\rangle \nonumber\\
  && + 2\mathrm{Re} \iint_{-\infty}^t\!d\tau dt_2^\prime\! \int_{-\infty}^{t_2^\prime}\!dt_1^\prime \big\langle \psi(-\!\infty) \big|\mathcal{U}_0(-\infty, t_1^\prime)  \mathcal{H}_{\rm pr}^{\rm (ab)\dagger}(t_1^\prime)\mathcal{U}_0(t_1^\prime,t_2^\prime)\mathcal{H}_{\rm pr}^{\rm (ab)\dagger}(t_2^\prime) \mathcal{U}_0(t_2^\prime,t) a_{\qbf_i\alpha}a^\dagger_{\qbf_s\beta}a_{\qbf_s\beta}a^\dagger_{\qbf_i\alpha}\mathcal{U}_0(t,\tau)\nonumber\\
  && \mathcal{H}_{\rm pr}^{\rm (sc)}(\tau) \mathcal{U}_0(\tau,-\infty) \big|\psi(-\!\infty) \big\rangle 
  + \iint_{-\infty}^t\!dt_2dt_2^\prime\! \int_{-\infty}^{t_2^\prime}\!dt_1^\prime\!\int_{-\infty}^{t_2}\!dt_1 \big\langle \psi(-\!\infty) \big|\mathcal{U}_0(-\infty, t_1^\prime) \mathcal{H}_{\rm pr}^{\rm (ab)\dagger}(t_1^\prime)\mathcal{U}_0(t_1^\prime,t_2^\prime)\mathcal{H}_{\rm pr}^{\rm (ab)}(t_2^\prime)\nonumber\\ 
  &&   \mathcal{U}_0(t_2^\prime,t) a_{\qbf_i\alpha}a^\dagger_{\qbf_s\beta}a_{\qbf_s\beta}a^\dagger_{\qbf_i\alpha}\mathcal{U}_0(t,t_2)\mathcal{H}_{\rm pr}^{\rm (ab)\dagger}(t_2) \mathcal{U}_0(t_2,t_1)  \mathcal{H}_{\rm pr}^{\rm (ab)}(t_1) \mathcal{U}_0(t_1,-\infty) \big|\psi(-\infty) \big\rangle.
\end{eqnarray}
\end{widetext}
The first term contributes only to the elastic background. The last term involving intermediate states between $t_1$ and $t_2$ is related to ``resonant" scattering. In contrast, the second and third terms are associated with ``nonresonant" and ``mixed" scatterings, respectively. Like in the equilibrium case, when the incident photon frequency $\omega_i$ is off resonance to any excited state, the last two terms in Eq.~\eqref{eq:crossSec1} can be ignored. Moreover, in the optical limit $\qbf \approx 0$, the odd parity of $\hat{j}_\alpha(0)$ forbids any finite resonant contribution. Without resonant intermediate states, the absolute energies of incident and scattering photons are irrelevant; only the energy difference $\omega = \omega_i-\omega_s$ is important for nonresonant scattering. Therefore, the time-resolved nonresonant optical Raman cross section can be written as 
\begin{eqnarray}\label{eq:crossSec2}
R_{\alpha\!\beta}(\omega,\!t)\! &=&\! 4\!\iint\! d\tau d\tau^\prime\!\big\langle \mathcal{U}_0(-\infty, \tau^\prime)  \mathcal{H}_{\rm pr}^{\rm (sc)\dagger}(\tau^\prime)\mathcal{U}_0(\tau^\prime,\infty)\nonumber\\
&&a_{\qbf_i\!\alpha}a^\dagger_{\qbf_s\!\beta}a_{\qbf_s\!\beta}a^\dagger_{\qbf_i\!\alpha}
\mathcal{U}_0(\infty,\tau)\mathcal{H}_{\rm pr}^{\rm (sc)}(\tau)\mathcal{U}_0(\tau,-\infty)\big\rangle. \nonumber\\
\end{eqnarray}
 
As detectors collect response signals over a time period much longer than the probe pulse width, the integral limit can be set to $+\infty$. Here, $t$ in $R_{\alpha\beta}(\omega,t)$ indicates the center of the probe profile (which will be introduced later). Using Wick's theorem, Eq.~\eqref{eq:crossSec2} can be simplified to
\begin{eqnarray}\label{eq:crossSec3}
R_{\alpha\!\beta}(\omega,\!t)\! &=&\!\iint_{-\infty}^{\infty} dt_1 dt_2\,\chi_{\alpha\beta}(t_1, t_2)s_{\qbf_i}(\infty,\tau^\prime)^*s_{\qbf_s}(\infty,\tau^\prime)\nonumber\\
&&s_{\qbf_s}(\infty,\tau)^*s_{\qbf_i}(\infty,\tau),
\end{eqnarray}
where the response function 
\begin{eqnarray}
	\chi_{\alpha\beta}(t_1, t_2) = i \langle \psi(t_2) | \hat{\gamma}_{\alpha\beta}(t_2) \mathcal{U}_0(t_2,t_1) \hat{\gamma}_{\alpha\beta}(t_1) | \psi(t_1)\rangle
\end{eqnarray}
and $s_\qbf(t_2,t_1)\!=\! \langle \psi(t_2)| a_\qbf^\dagger \mathcal{U}_0(t_2,t_1) a_\qbf |\psi(t_1)\rangle$. In the semiclassical limit, the photon annihilation operator gives the square root of the instantaneous photon number. Thus, $s_\qbf(t_2,t_1)\!\approx\! \sqrt{[n^{\rm ph}_\qbf(t_2)+1]n^{\rm ph}_\qbf(t_1)}e^{-i\omega_\qbf (t_1-t_2)}$. In the finite-probe-width limit ${t_2\rightarrow\infty}$, $n^{\rm ph}_\qbf(t_2)=0$, so the photon part contributes an instantaneous shape function with a phase factor $s_{\qbf_s}(\infty,\tau)^*s_{\qbf_i}(\infty,\tau) \approx g(\tau;t) e^{-i\omega \tau}$. Therefore,
\begin{eqnarray}\label{eq:crossSec4}
R_{\alpha\!\beta}(\omega,\!t)\! =\! \iint_{-\infty}^{\infty}\! dt_1 dt_2 e^{i\omega(t_2-t_1)}\!
g(t_1;\!t)g(t_2;\!t)\chi_{\alpha\beta}(t_1,\!t_2).
\end{eqnarray}
The probe shape function $g(t^\prime;t)$ can be approximated by a Gaussian pulse centered at time $t$ with width $\sigma_{\rm pr}$:
\begin{eqnarray}\label{eq:probeProfile}
g(t^\prime;t) = \frac1{\sqrt{2\pi}\sigma_{\rm pr}} e^{-(t^\prime-t)^2/2\sigma_{\rm pr}^2}.
\end{eqnarray}

The polarization can be decomposed into an irreducible representation of $D_{4h}$ point group in the long-wavelength limit. Therefore, Eq.~\eqref{eq:crossSec4} produces the time-resolved nonresonant optical Raman cross section by replacing $\chi_{\alpha\beta}(t_1, t_2)$ with various vertices. Specifically, in the $A_{1g}$ channel
\begin{eqnarray}\label{eq:noneqA1ggamma}
  \hat{\gamma}_{A_{1g}}\!(t) = t_h\! \sum_{\textbf{k}}\! \big[\cos \big(k_x\!-\!A_x\!(t)\big) +\cos \big(k_y\!-\!A_y\!(t)\big)\big]n_{\textbf{k}},
\end{eqnarray}
and in the $B_{1g}$ channel
\begin{eqnarray}\label{eq:noneqB1ggamma}
  \hat{\gamma}_{B_{1g}}\!(t) = t_h\! \sum_{\textbf{k}}\! \big[\cos \big(k_x\!-\!A_x\!(t)\big) -\cos \big(k_y\!-\!A_y\!(t)\big)\big]n_{\textbf{k}}.
\end{eqnarray}
Note that $\mathbf{A}(t)$ is the pump (instead of the probe) field, which should be treated explicitly in the calculation.

\subsection{Probe-Induced Linewidth Broadening}
When the pump field is turned off $\mathbf{A}(t)=0$, the scattering vertices Eqs.~\eqref{eq:noneqA1ggamma} and \eqref{eq:noneqB1ggamma} are identical to Eqs.~\eqref{eq:eqA1ggamma} and \eqref{eq:eqB1ggamma}; the Hamiltonian $\Ham_0(t)$ becomes time-independent, and the system is time-translationally invariant: $\chi_{\alpha\beta}(t_1, t_2) = \chi_{\alpha\beta}(t_2-t_1)$. In this case, the time-dependent Raman cross section Eq.~\eqref{eq:crossSec4} simplifies to
 \begin{eqnarray}\label{eq:crossSec5}
R_{\alpha\beta}(\omega,t) &=&\!\frac1{2\pi\sigma_{\rm pr}^2}\ \iint_{-\infty}^{\infty}\! dt_1 dt_2\,e^{i\omega(t_2-t_1)}  e^{-(t_1-t)^2/2\sigma_{\rm pr}^2}\nonumber\\
&&\times e^{-(t_2-t)^2/2\sigma_{\rm pr}^2}
\chi_{\alpha\beta}(t_2-t_1)\nonumber\\
&=& \frac1{2\pi\sigma_{\rm pr}^2}\! \iint_{-\infty}^{\infty}\! dT d\tau\,e^{i\omega\tau} \! e^{-\frac{(T-t)^2}{\sigma_{\rm pr}^2}}e^{-\frac{\tau^2}{4\sigma_{\rm pr}^2}}
\chi_{\alpha\beta}(\tau)\nonumber\\
&=& \frac{\sigma_{\rm pr}^2}{\pi} \int_{-\infty}^{\infty}d\omega^\prime\,  e^{-(\omega^\prime-\omega)^2\sigma_{\rm pr}^2}
\mathcal{R}_{\alpha\beta}(\omega^\prime),
\end{eqnarray}
where $\mathcal{R}_{\alpha\beta}(\omega^\prime)$ is nothing but the equilibrium Raman cross section Eq.~\eqref{chp2:nonResRaman} with a zero (Lorentzian) broadening, $\delta =0$. Therefore, the nonequilibrium Raman response $R_{\alpha\beta}(\omega,t)$ in the zero-pump limit reproduces exactly the equilibrium one with a (Gaussian) linewidth $\sim 1/(\sqrt{2}\sigma_{\rm pr})$. In fact, if the probe shape function is set as $g(t^\prime;t) \sim e^{-\delta|t-t^\prime|}$, the equilibrium cross section Eq.~\eqref{chp2:nonResRaman} can be exactly recovered. To mimic a realistic probe, however, the Gaussian shape function Eq.~\eqref{eq:probeProfile} is more appropriate and is adopted in this paper. The finite probe duration causes a finite energy broadening, which may lead to (limited) uncertainty of physical observables, such as the effective temperature discussed in Sec.~\ref{sec:effectiveT}.

\section{Numerical Time-Resolved Raman Spectra on a Correlated System}\label{sec:result}
With the above formalism, below we use exact diagonalization to compute the time-resolved Raman spectra on the square-lattice single-band Hubbard model:
\begin{eqnarray}\label{eq:Hubbard}
  \Ham = -t_h\sum_{\langle \ibf,\jbf\rangle,\sigma} c_{\ibf\sigma}^\dagger c_{\jbf\sigma} + U\sum_\ibf n_{\ibf\uparrow} n_{\ibf\downarrow}.
\end{eqnarray}
Without further specification, the Hubbard interaction is set to $U=8t_h$ as a typical choice for high-$T_c$ cuprate compounds. This choice leads to an effective spin exchange energy $J=4t_h^2/U = 0.5t_h$. The material-specific value of $t_h$ is  $\sim 300-400$ meV for the cuprates. We consider only N.N. hopping, and the ground state of the model is a Mott insulator with a predominant antiferromagnetic order.

\begin{figure}[!ht]
\begin{center}
\includegraphics[width=\columnwidth]{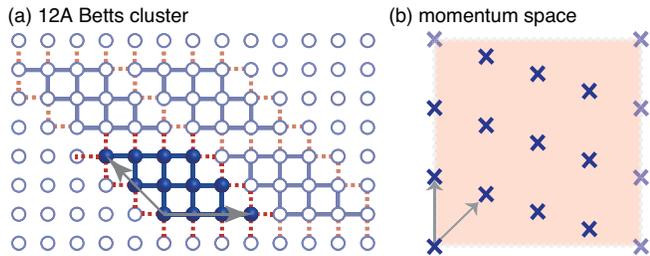}
\caption{\label{fig:s0} 
The Betts 12A cluster in (a) real space and (b) momentum space. The gray arrows denote basis vectors. The solid and dotted lines in (a) represent respectively the intra- and inter-cluster hopping terms.
}
\end{center}
\end{figure}

As discussed above, while the probe field $\textbf{A}^{\rm(pr)}$ is treated with perturbation theory, the pump field $\mathbf{A}(t)$ is considered explicitly through a Peierls substitution. Here we use an oscillatory Gaussian vector potential in the temporal gauge to simulate a pulsed laser pump:
\begin{eqnarray}
\mathbf{A}(t)=A_0 e^{-t^2/2\sigma^2}\cos(\Omega t)\, \mathbf{e}_{\rm pol},
\end{eqnarray}
where $A_0$, $\sigma$, $\Omega$, and $\mathbf{e}_{\rm pol}$ are respectively the pump amplitude, width, frequency, and polarization. The time $t=0$ corresponds to the center of the pump. The calculation is performed on the Betts 12A cluster with periodic boundary conditions. Due to the cluster's tilted geometry, the diagonal polarization in momentum space in fact reflects the horizontal polarization in real space [see Fig.~\ref{fig:s0}]. We use the parallel Arnoldi method\cite{lehoucq1998arpack,  jia2017paradeisos} to determine the equilibrium ground-state wavefunction $\big|\psi(t=-\infty)\big\rangle$, and the Krylov subspace technique\cite{manmana2007strongly, balzer2012krylov, hochbruck1997krylov} to evaluate the wavefunction's time evolution $|\psi(t\!+\!\delta t)\rangle\!=\!e^{-i\mathcal{H}(t)\delta t}|\psi(t)\rangle$.

Below we first give an overview of the main features of time-resolved Raman spectra in the $B_{1g}$ channel. We then analyze two important processes: the Floquet renormalization of spin exchange and effective thermalization, each of which can dominate at different pump conditions. At the end of this section, we provide a comprehensive discussion on the impact of pump polarization and probe width on the nonequilibrium Raman spectra.

\subsection{Time-Resolved $B_{1g}$ Raman Spectra}
\begin{figure}[!th]
\includegraphics[width=\columnwidth]{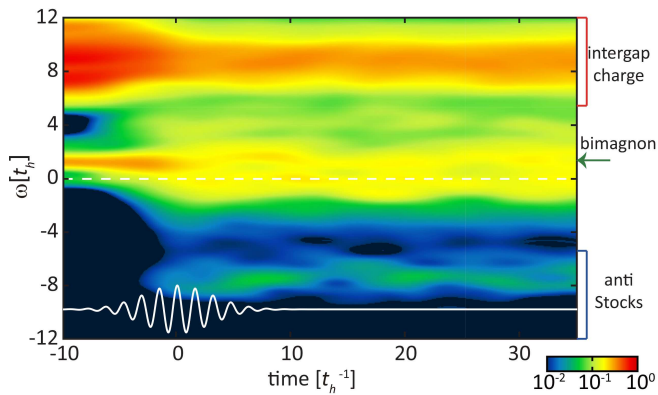}
\caption{\label{fig:ramanspec}
Time-resolved $B_{1g}$ Raman spectra with pump polarization $\mathbf{e}_{\rm pol} =\mathbf{e}_{\rm x}$, frequency $\Omega=4t_h$, and amplitude $A_0=0.6$. The bimagnon and Stokes/anti-Stokes charge excitations are marked on the right to guide the eye.
Energy zero (defined as the equilibrium ground-state energy) is denoted by the dashed white line.
The oscillatory Gaussian pump is drawn as a solid white curve.
}
\end{figure}

\begin{figure}[!t]
\begin{center}
\includegraphics[width=\columnwidth]{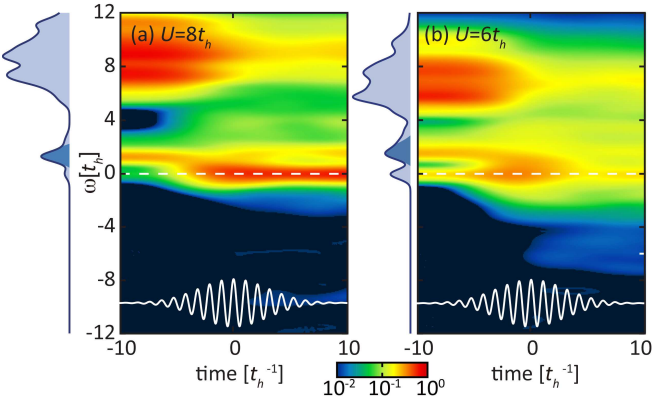}
\caption{\label{fig:s2}
Time-resolved $B_{1g}$ Raman spectra with $\mathbf{e}_{\rm pol} =\mathbf{e}_{\rm x}$, $\Omega=6t_h$, and $A_0=0.3$ for (a) $U=8t_h$ and (b) $U=6t_h$. The plots are drawn in the same manner as in Fig.~\ref{fig:ramanspec}. The shaded curves to the left of each panel show the corresponding equilibrium spectra; the bimagnon peak is denoted by a darker color.
}
\end{center}
\end{figure}

Raman spectra usually exhibit a prominent elastic signal, and the $B_{1g}$ channel is usually adopted to resolve the features of low-energy excitations. Figure \ref{fig:ramanspec} shows the time-resolved $B_{1g}$ Raman spectra with the horizontal pump polarization $\mathbf{e}_{\rm pol} =\mathbf{e}_{\rm x}$. The pump frequency and amplitude are set to $\Omega=4t_h$ and $A_0=0.6$, respectively. Before the pump enters, the equilibrium spectrum exhibits a low-energy peak at $\sim 1.3 t_h$ attributed to bimagnon excitation. In the strong-coupling limit $U \rightarrow \infty$, the bimagnon energy of $\sim 3J$ represents two locally bounded spin-flip excitations. The excitation energy is further reduced due to finite charge fluctuations. Further calculations with different strengths of $U$ have supported the assignment of the low-energy bimagnon peak [discussed latter in Fig.~\ref{fig:s2}]. In addition, a cloud of cross-gap charge excitations exists above the Mott gap, approximately within the energy range $[U-4t_h, U+4t_h]$. With our current choice of $U$, these charge modes are well-separated from the bimagnon peak, which thereby provides an opportunity to track these excitations individually. Since the 12A cluster breaks $C_4$ symmetry, the ground state shows a small, unphysical elastic peak. No signals are observed below zero energy, as the system is at the ground state.

In the presence of the pump, the bimagnon energy softens transiently and becomes indistinguishable from the elastic peak. This can be attributed to a renormalized spin exchange interaction through the Floquet photoassisted process, as discussed later in Eqs. (\ref{eq:Floquetrenormalization}) and (\ref{eq:bimagnonFloquet}). Meanwhile, the anti-Stokes features start to appear with the pump, and the Stokes excitations across the Mott gap are suppressed accordingly. This is a signature of pump-induced thermalization. Moreover, the energies of cross-gap excitations are modified by the pump, and new spectral poles near $4t_h$ start to develop. These new ``in-gap" states indicate the appearance of many-body excitations beyond a simple heating effect. As mentioned before, the three ultrafast processes in Fig.~\ref{fig:cartoon} are all reflected in the time-resolved Raman spectra.

Figure~\ref{fig:s2} examines the time-resolved $B_{1g}$ Raman spectra for two different strengths of Hubbard $U$. As the spin exchange energy is roughly $J = 4t_h^2/U$, the bimagnon energy for $U=8t_h$ is smaller than that for $U=6t_h$, while the charge gap is larger in the former by definition. When the pump effect is present, the bimagnon softening is more obvious for $U=6t_h$, since the electron is more delocalized compared to that for $U=8t_h$. In general, the $U=6t_h$ case is more vulnerable to the same pump condition, with greater bigmagnon softening, more obvious anti-Stokes features, and stronger spectral redistribution inside the charge gap. A more delocalized system also makes the calculation more sensitive to the small-cluster size/geometry and causes a stronger equilibrium elastic peak. These qualitative trends of the Raman spectra further support the assignments of different spectral features and corresponding physical processes. Note that in Fig.~\ref{fig:s2} we consider a pump frequency $\Omega=6t_h$, which should be on resonance with certain excited states in both $U=6t_h$ and $U=8t_h$. Due to the expected strong thermalization and many-body effects, we thereby select a relatively weak pump strength $A_0=0.3$.

\subsection{Floquet Manipulation of Bimagnon Excitation}

\begin{figure}
\includegraphics[width=\columnwidth]{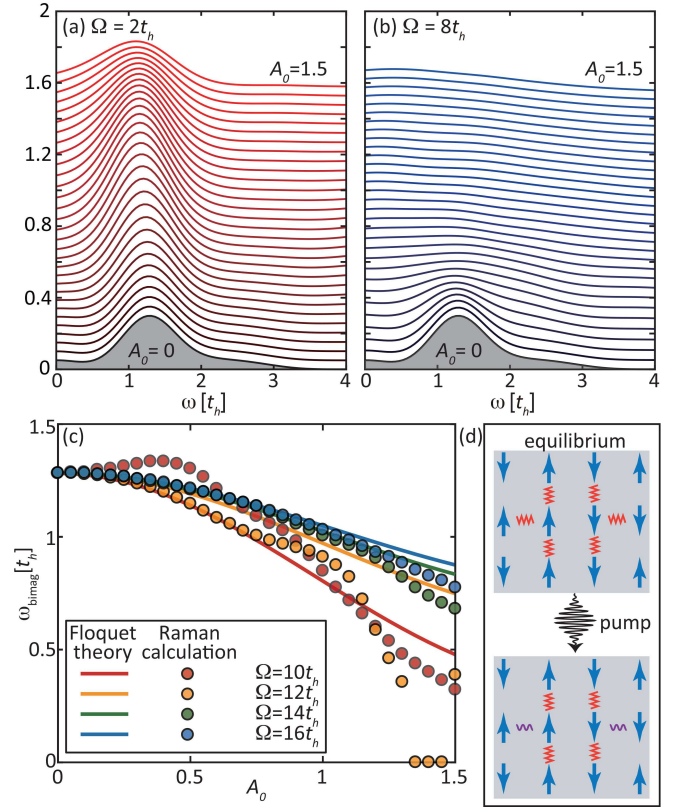}
\caption{\label{fig:EDC}
The $B_{1g}$ Raman spectra with pump amplitude $A_0$ varying from 0 (bottom) to $1.5$ (top) at pump frequency (a) $\Omega=2t_h$ and (b) $\Omega=8t_h$. The time is fixed at $t=0$ (the center of the pump). (c) Bimagnon energies as a function of $A_0$ under high-frequency pumps ($\Omega > U$). The solid curves are predicted by the Floquet theory in the nonresonant limit. (d) Schematic cartoon showing specific spin exchange bonds altered by a pump pulse field.}
\end{figure}

Due to both thermalization and many-body scattering, the bimagnon excitation can become less well-defined. Figure~\ref{fig:EDC} tracks the Raman response for the bimagnon peak at time $t=0$ under various pump profiles. At a small pump frequency $\Omega = 2t_h$ [Fig.~\ref{fig:EDC}(a)], the bimagnon energy remains unchanged, but its peak width is gradually broadened with increasing pump amplitude. This is consistent with the thermalization mechanism, where a small number of particle-hole excitations are created across the Mott gap. On the other hand, at a larger pump frequency $\Omega = 8t_h$ [Fig.~\ref{fig:EDC}(b)], the bimagnon energy and width can strongly depend on the pump amplitude. The non-thermal mechanisms underlying these changes are discussed below.

As shown in Fig.~\ref{fig:EDC}(c) for $\Omega=12t_h - 16 t_h$, a high-frequency pump with strong amplitude can soften significantly the spin exchange $J$ or the bimagnon energy. Here, in order to reduce the impact of the elastic mode and better resolve the bimagnon peak, we consider the $R_{B_{1g}}(\omega,t) - R_{B_{1g}}(-\omega,t)$ spectra to perform the peak analysis. This softening behavior can be understood by the Floquet renormalization at the nonresonant limit ($m\Omega \neq U$)\cite{mentink2015ultrafast}:
\begin{equation}\label{eq:Floquetrenormalization}
    \frac{J(A_0)}{J(A_0=0)}=\sum_{m=-\infty}^{+\infty}\frac{\mathcal{J}_{|m|}(A_0)^2}{1+m\Omega/U},
\end{equation}
in which $\mathcal{J}_m(x)$ is the Bessel function of the first kind. With our choice of the pump polarization, one third of the bonds participating in bimagnon excitation will be strongly altered, as illustrated in Fig.~\ref{fig:EDC}(d). Therefore, the resulting dynamically renormalized bimagnon energy is
\begin{equation}\label{eq:bimagnonFloquet}
     \omega_{\rm bimag}(A_0) \approx \omega_{\rm bimag}(0)\left[\frac23 + \frac13\sum_{m=-\infty}^{+\infty}\frac{\mathcal{J}_{|m|}(A_0)^2}{1+m\Omega/U}\right].
\end{equation}
Figure~\ref{fig:EDC}(c) shows that the Floquet theory indeed can capture the softening of the bimagnon. On the other hand, since a finite-width pump contains all frequency components, it cannot be completely off resonance. When the pump strength is strong enough, the Floquet theory prediction can deviate from the Raman calculation, as shown in Fig.~\ref{fig:EDC}(c). This deviation is more apparent at lower frequency ($\Omega=10$ and $12t_h$) than at higher frequency ($\Omega=16t_h$), since the former is closer to $U$.

When the pump frequency is on resonance, the magnon softening can be accompanied by other effects. At $\Omega = 8 t_h = U$, shown in Fig.~\ref{fig:EDC}(b), the bimagnon in fact first hardens with increasing $A_0$, and the energy also deviates from the Floquet prediction. Similar behaviors are also seen at $\Omega=10t_h$ shown in Fig.~\ref{fig:EDC}(c). This deviation signals a coherent many-body renormalization due to the draining of electrons to unoccupied, real states. These occupancies typically enter through the corresponding energy and momentum positions of the Floquet virtual states, but become heavily renormalized by many-body scattering\cite{wang2017producing}. The selected occupied states then reversely correct the effective interaction and spin exchange energy. The results in Fig.~\ref{fig:EDC} demonstrate the possibility of using specifically tailored pump frequency and amplitude to engineer the spin exchange interaction out of equilibrium in a well-controlled manner. In contrast to the previously predicted ultrafast control of spin exchange interaction using theoretical observables\cite{mentink2014ultrafast, mentink2015ultrafast}, the time-resolved Raman spectrum provides a practical approach to measure this change in a condensed-matter experiment.

\subsection{Extraction of Effective Temperature}\label{sec:effectiveT}

\begin{figure}[!b]
\includegraphics[width=\columnwidth]{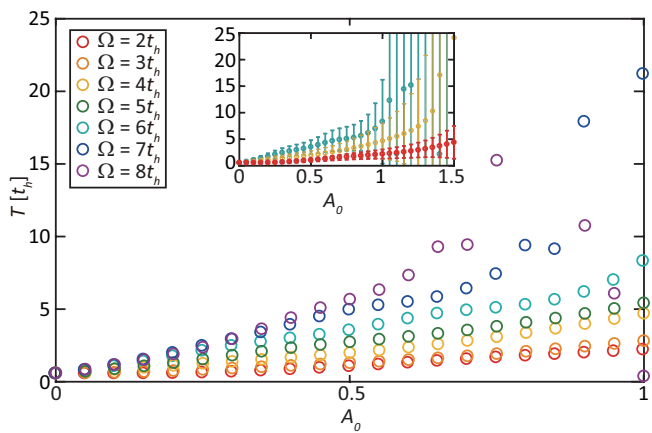}
\caption{\label{fig:temp}
Effective temperatures ${T_{\rm eff}}$ extracted from the post-pump ($t=10t_h^{-1}$) ratio of Stokes/anti-Stokes responses averaged over the energy range $4t_h <\omega< 12t_h$ for the $B_{1g}$ Raman spectra. The colors denote different pump frequencies $\Omega$. The inset shows mean values and standard deviations of ${T_{\rm eff}}$ for $\Omega=2$, 4 and 6$t_h$.}
\end{figure}

In addition to an energy shift, the bimagnon peak also broadens rapidly with increasing $A_0$ at high-frequency pumps. This broadening phenomenon is especially apparent under the resonance condition $\Omega \sim U$ [see for example Fig.~\ref{fig:EDC}(b)].  In the following we quantitatively analyze the Stokes/anti-Stokes responses after the pump (at time $t=10t^{-1}_h$) to show the clear distinction between on and off resonances. In the fluctuation-dissipation theorem, the structure factor can be written as 
\begin{eqnarray}
	\mathcal{R}(\omega) = \frac{1}{\pi}\frac{\imag[\chi(\omega)]}{e^{-\omega/T} - 1}.
\end{eqnarray}
Since the imaginary part of the response $\chi(\omega)$ is an odd function, ${\mathcal{R}(-\omega)}/{\mathcal{R}(\omega)} = e^{-\omega/T}$. Therefore, an effective temperature can be defined as
\begin{eqnarray}
T_{\rm eff}  (\omega) = \frac{\omega}{\ln \mathcal{R}(\omega)- \ln \mathcal{R}(-\omega)}.
\end{eqnarray}
Below we extract the effective temperatures $T_{\rm eff}$ averaged over the energy range $4t_h\!<\!\omega\!<\!12t_h$ for the $B_{1g}$ Raman spectra at each time. The uncertainty associated with the standard deviation then can be defined for each $T_{\rm eff}$. Within this procedure, $T_{\rm eff}$ remains nonzero even at $A_0=0$. This is because the fluctuation-dissipation theorem is exact only for spectra with zero linewidth. Therefore, our finite-width probe would give rise to a small error of $1/(\sqrt{2}\sigma_{\rm pr})=0.3t_h$ in estimating $T_{\rm eff}$.

\begin{figure}[!t]
\begin{center}
\includegraphics[width=\columnwidth]{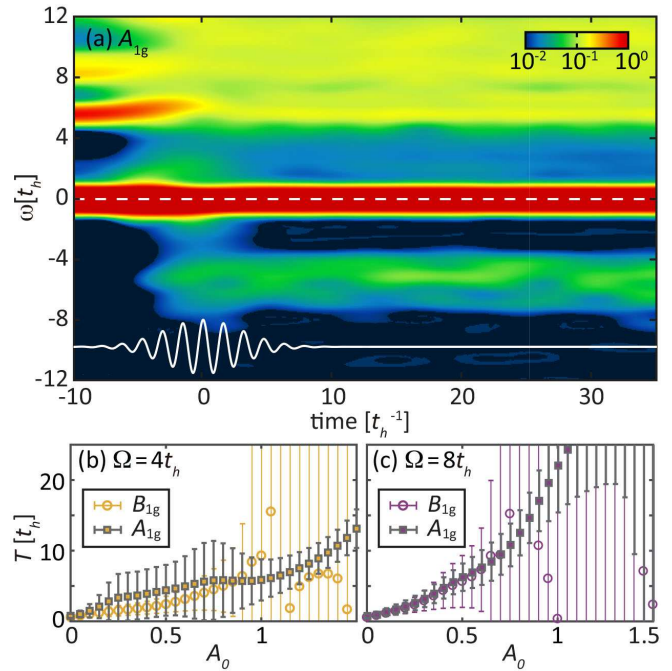}
\caption{\label{fig:s5} 
(a) Time-resolved $A_{1g}$ Raman spectra with $\mathbf{e}_{\rm pol} =\mathbf{e}_{\rm x}$, $\Omega=4t_h$, and $A_0=0.6$. The pump profile is identical to that in Fig.~\ref{fig:ramanspec}. The plot is drawn in the same manner as in Fig.~\ref{fig:ramanspec}. (b, c) Comparison of effective temperatures ${T_{\rm eff}}$ extracted from $A_{\rm 1g}$ and $B_{\rm 1g}$ Raman spectra at a post-pump time $t=10t_h^{-1}$ under different pump frequencies, using the ratio of Stokes/anti-Stokes responses averaged over the energy range $4t_h <\omega< 12t_h$.
}
\end{center}
\end{figure}

Figure~\ref{fig:temp} shows that increasing $\Omega$ would enhance $T_{\rm eff}$ linearly at small $A_0$, as the fluence is roughly proportional to $\Omega A_0^2$. However, at larger pump amplitudes $A_0 \gtrsim 1$, the uncertainty of $T_{\rm eff}$ can become comparable to its mean or even diverge [see the inset of Fig.~\ref{fig:temp}]. This indicates that many-body renormalization becomes dominant over thermalization under high-frequency and large-amplitude pumps. Many-body scattering can significantly alter the wavefunction and lead to a nonequilibrium state violating the fluctuation-dissipation theorem. Reflected in the Raman spectra, it is the break down of extracting effective temperatures from the Stokes/anti-Stokes responses under strong pump pulse fields.

To further study the effective thermalization picture, we also compute the time-resolved Raman spectra in the $A_{1g}$ channel. As shown in Fig.~\ref{fig:s5}(a), the strong elastic peak at $\omega=0$ is the predominant feature. Due to this strong elastic signal, the bimagnon peak cannot be resolved, but the cross-gap charge excitations at $\omega>4t_h$ are still clearly visible. After the pump, a number of anti-Stokes responses appear. We then obtain the corresponding $T_{\rm eff}$ by following the same procedure as in Fig.~\ref{fig:temp}. At $\Omega = 4t_h$ [Fig.~\ref{fig:s5}(b)], the effective temperatures extracted from both the $A_{1g}$ and $B_{1g}$ channels match fairly well within the errors for $A_0 \lesssim 0.9$. Due to the strong elastic peak, the $A_{1g}$ spectrum overestimates $T_{\rm eff}$ and contains larger errors for small $A_0$. The consistency in $T_{\rm eff}$ extracted from two different Raman channels confirms that effective heating of electrons is a reasonable description of the intrinsic nonequilibrium physics for small $A_0$. The same conclusion is also reached for $\Omega=8t_h$ [Fig.~\ref{fig:s5}(c)], where the deviation does not occur until $A_0=0.8$.  Due to the larger fluence of a higher-frequency pump, it is expected that the thermalization picture breaks down at smaller $A_0$. Finally, we note that the concept of effective heating may seem to work better in the $A_{1g}$ channel [see for example Fig.~\ref{fig:s5}(b), where the error bar does not diverge even at $A_0 = 1.5$]. This is because different cross-gap charge modes can be selectively excited by different scattering vertices, so a comprehensive examination of different Raman channels may be necessary for the thermalization description.

Inclusion of next-nearest-neighbor hopping $t_h^\prime$ in the Hamiltonian [Eq.~\eqref{eq:Hubbard}] will lead to a non-zero $B_{\rm 2g}$ Raman spectrum, which also can be employed to extract $T_{\rm eff}$. Using the common choice of $t_h^\prime = -0.3t_h$ for the cuprates, we find that the time-resolved $B_{\rm 2g}$ response [not shown] is two orders of magnitude smaller than the $B_{\rm 1g}$ and $A_{\rm 1g}$ spectra. Due to the weak effective next-nearest-neighbor spin exchange $J' = (t^\prime_h/t_h)^2 J\sim 0.1J$ and the strong antiferromagnetism in a half-filled Hubbard model, the bimagnon signal in the $B_{\rm 1g}$ channel does not exhibit a noticeable difference between $t_h^\prime = -0.3t_h$ and $t_h^\prime = 0$.

\begin{figure}[!t]
\begin{center}
\includegraphics[width=\columnwidth]{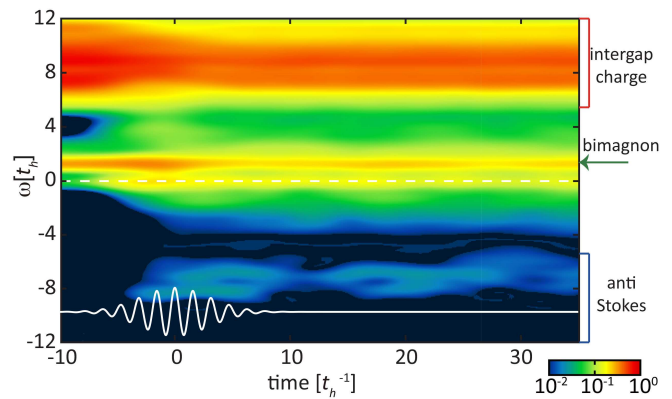}
\caption{\label{fig:s1} 
Time-resolved $B_{1g}$ Raman spectra with $\Omega=4t_h$ and $A_0=0.6$, for the tilted polarization corresponding to the horizontal momentum-space direction. The plot is drawn in the same manner as in Fig.~\ref{fig:ramanspec}.
}
\end{center}
\end{figure}

\subsection{Impact of Pump and Probe Conditions}
So far, we have considered only the horizontal pump polarization $\mathbf{e}_{\rm pol} =\mathbf{e}_{\rm x}$, which corresponds to the diagonal direction in momentum space, due to the tilted geometry of the Betts 12A cluster [see Fig.~\ref{fig:s0}]. Changing $\mathbf{e}_{\rm pol}$ to a tilted polarization that corresponds to the horizontal momentum-space direction would result in a projection along both basis vectors. With such a projection, the renormalization of spin exchange energy is relatively minor, as shown in Fig.~\ref{fig:s1}. This is because the horizontal momentum-space direction is less nested, and the pump field does not help cross-gap excitations much. Therefore, the thermalization effect is also less obvious.

\begin{figure}[!t]
\begin{center}
\includegraphics[width=8cm]{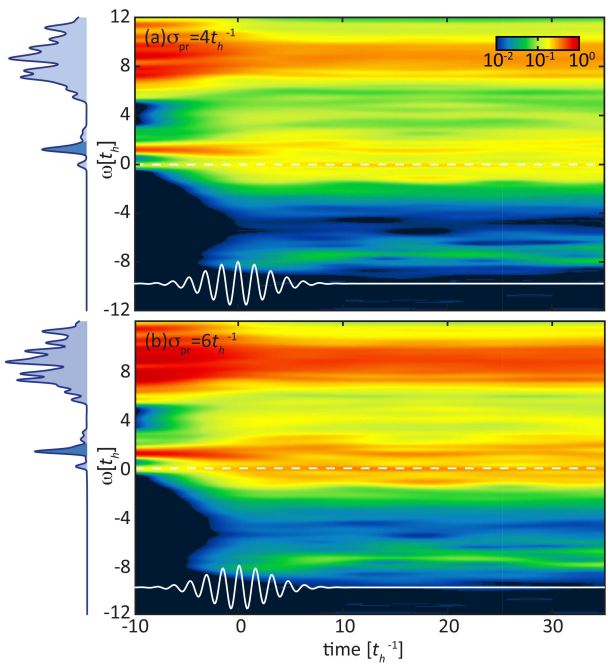}
\caption{\label{fig:s3} 
Time-resolved $B_{1g}$ Raman spectra with probe pulse width (a) $\sigma_{\rm pr}=4t_h$ and (b) $\sigma_{\rm pr}=6t_h$. The pump profile is identical to that in Fig.~\ref{fig:ramanspec}. The plots are drawn in the same manner as in Fig.~\ref{fig:ramanspec}. 
}
\end{center}
\end{figure}

While the pump profile changes the dynamics of the ultrafast processes discussed above, the probe profile determines only the spectral resolution in both frequency and time domain, without changing the underlying physics. Figure \ref{fig:s3} shows the time-resolved $B_{1g}$ spectra probed by different pulse widths $\sigma_{\rm pr}=4t_h^{-1}$ and $\sigma_{\rm pr}=6t_h^{-1}$. The pump profile is the same as that in Fig.~\ref{fig:ramanspec} (which has $\sigma_{\rm pr}=2t_h^{-1}$). Due to the uncertainty principle, a wider probe has less time but more energy/frequency resolutions, which is reflected already in the equilibrium spectra. The finer frequency structure with wider probe width in Fig.~\ref{fig:s3} shows that the bimagnon peak is a sharp, well-defined quasiparticle. In addition, the continuum of excitation above the charge gap consists of many poles, which can be more easily resolved with a wider probe. Figure \ref{fig:s3} also shows that both thermalization and Floquet renormalization happen during the pump. After the pump, a wider probe clearly resolves the softening and broadening of the bimagnon peak. On the other hand, albeit with a gain in frequency resolution, a wider probe has a bad time resolution. For example, unlike the oscillatory features observed in Fig.~\ref{fig:s3}(a), the $\sigma_{\rm pr}=6t_h^{-1}$ spectrum in Fig. \ref{fig:s3}(b) exhibits almost a constant structure in time after the pump. This constant structure is essentially the time average of that in a narrower probe.

Finally, we connect our theory results to real physical units. For $t_h = 350$ meV in a cuprate material, the corresponding time-scale is roughly $t_h^{-1}\approx 11.82$ fs. Therefore, the duration of the pump pulse considered here is 213 fs. The pump frequency discussed in this section varies from 2$t_h$ to 16$t_h$, which corresponds to a photon energy between $0.7$ and $5.6$\,eV. For a pump amplitude $A_0 = 0.5$, the corresponding fluence is $0.007\,J/$cm$^2$ for $\Omega=2t_h$ and $0.46\,J/$cm$^2$ for $\Omega=16\,t_h$. The clear Floquet renormalization happens at the ultraviolet end, while typical thermalization occurs at the infrared end. Therefore, to achieve a faithful Floquet renormalization of bimagnon as in Fig.~\ref{fig:EDC}(c), one should employ a near-ultraviolet laser with a fluence less than $0.5\,J/$cm$^2$. To extract a well-defined effective temperature as in Fig.~\ref{fig:temp}, one could employ a weak infrared or even terahertz pump, without much of a requirement for the monochromaticity. In the latter case, one should pay particular attention to the shift of $T_{\rm eff}$ induced by the finite-probe width, as mentioned in Sec.~\ref{sec:effectiveT}.

\section{Conclusion}\label{sec:conclusion}
In summary, we have derived the theory of time-resolved Raman scattering and evaluated the $A_{1g}$ and $B_{1g}$ Raman spectra on a pumped square-lattice single-band Hubbard model. The spectra were shown to exhibit different ultrafast processes, and each of them can become dominant under different pump conditions. In particular, thermalization dominates at small pump frequencies, and an effective temperature can be extracted. In contrast, for large-frequency ``off resonance'' pumps, the Floquet theory successfully captures the renormalization of effective spin exchange interaction as manifested in the softening of the bimagnon energy. When the pump frequency is ``on resonance" with the Mott gap, coherent many-body effects start to contribute. While thermalization still dominates at low pump fluences, many-body scattering takes over at large pump amplitudes and results in nonequilibrium states violating the Fermi-Dirac distribution. Time-resolved Raman scattering thereby provides a platform for exploring different ultrafast processes in a pump-probe experiment. With tailored pump conditions, it also opens up new opportunities to directly probe and engineer spin exchange interaction out of equilibrium. In accordance with our theoretical predictions, detailed experimental investigations of the pump amplitude, frequency, and polarization would be intriguing future studies, especially for strongly correlated systems.

\section*{Acknowledgments}
Y.W. is supported by the Postdoctoral Fellowship in Quantum Science of the Harvard-MPQ Center for Quantum Optics and AFOSR-MURI Quantum Phases of Matter (Grant No.~FA9550-14-1-0035). T.P.D. acknowledges support from the U.S. Department of Energy, Office of Science, Office of Basic Energy Sciences, Division of Materials Sciences and Engineering, under Contract No.~DE-AC02-76SF00515.  C.-C. C. is supported in part by the National Science Foundation under Grant No.~OIA-1738698. This research used resources of the National Energy Research Scientific Computing Center (NERSC), a U.S.~Department of Energy Office of Science User Facility operated under Contract No.~DE-AC02-05CH11231.

\bibliography{paper}
\end{document}